\documentclass[a4paper,11pt]{article}

\usepackage{jheppub} 

\makeatletter
\gdef\@fpheader{}
\makeatother

\usepackage[T1]{fontenc} 

\usepackage{verbatim}

\DeclareMathOperator{\sgn}{sgn}
\usepackage[normalem]{ulem}
\usepackage{soul}
\def\lsim{\mathrel{\rlap{\lower4pt\hbox{\hskip1pt$\sim$}}
    \raise1pt\hbox{$<$}}} 
\def\gsim{\mathrel{\rlap{\lower4pt\hbox{\hskip1pt$\sim$}}
    \raise1pt\hbox{$>$}}} 

\newcommand{\RN}[1]{%
	\textup{\uppercase\expandafter{\romannumeral#1}}%
}

\title{\boldmath Spontaneous Leptogenesis in Higgs Inflation}

\author{Sung Mook Lee,\footnote{E-mail: \tt sungmook.lee@yonsei.ac.kr}}
\author{Kin-ya Oda,\footnote{E-mail: \tt odakin@phys.sci.osaka-u.ac.jp}}
\author{and Seong Chan Park \footnote{E-mail: \tt sc.park@yonsei.ac.kr}}

\affiliation[1,3]{Department of Physics and IPAP, Yonsei University, Seoul 03722, Republic of Korea}
\affiliation[2]{Department of Physics, Osaka University, Osaka 560-0043, Japan}

\abstract{
We propose a scenario of spontaneous leptogenesis in Higgs inflation with help from two additional operators: the Weinberg operator (Dim 5) and 
the derivative coupling of the Higgs field and the current of  lepton number (Dim 6). The former is responsible for lepton number violation and the latter induces chemical potential for lepton number.  The period of rapidly changing Higgs field, naturally realized in Higgs inflation during the reheating, allows large enhancement in the produced asymmetry in lepton number, which is eventually converted into baryon asymmetry of the universe. This scenario is compatible with high reheating temperature of Higgs inflation model.
}

\begin{document} 
\maketitle
\flushbottom

\section{Introduction}
Mysteriously the universe is populated by matter rather than anti-matter. This matter-antimatter asymmetry has been a long-standing problem in  particle physics and cosmology since it could not be understood within the framework of the Standard Model (SM). Conventional wisdom is to look for a theory which satisfies the Sakharov conditions when CPT symmetry is assumed~\cite{Sakharov:1967dj}: (1) Baryon number violation, (2) C and CP violation, and (3) out of equilibrium. Even though all the conditions are fulfilled in the SM, the CP violation is too weak to account for the observed amount of asymmetry. The asymmetry is conveniently represented by the difference between the number densities of baryons and anti-baryons, $n_B = n_b -n_{\bar b}$,\footnote{We will use small letters $ b(\overline{b}) $ and $ \ell(\overline{\ell}) $ for particle (anti-particle) numbers density, and capital letters $ B $ and $ L $ for `total' number density. Therefore, $ n_{B} = n_{b} - n_{\overline{b}} $ and $ n_{L} = n_{\ell} - n_{\overline{\ell}} $. } and the number density of photons, $n_\gamma$ \cite{Zyla:2020zbs}:
\begin{align}
\eta_{B} \equiv \frac{n_{B}}{n_{\gamma}} \simeq (6.12 \pm 0.04)\times 10^{-9}.
\end{align}

One of the most popular ideas is leptogenesis in a GUT theory~\cite{Fukugita:1986hr} (for review see e.g. \cite{Davidson:2008bu, Blanchet:2012bk, Bodeker:2020ghk}), where right handed neutrinos have Majorana masses that allow lepton number violation. The sphaleron processes eventually convert the lepton asymmetry to a baryon asymmetry~\cite{Khlebnikov:1988sr}.  However, the large Majorana mass is beyond the coverage of currently available or future coming experiments in near future so that the idea is hardly confirmed by observational data~\cite{Chun:2003ej}. For possible search of sphaleron effect, see Ref.~\cite{Jho:2018dvt}. 

When inflaton is involved, the issue is subtle since spatially homogeneous inflaton background in an expanding universe does not respect CPT symmetry: notably, the coupling between the inflaton and the divergence of the current of lepton (or baryon) number, $\partial_\mu j_L^\mu~(\partial_\mu j_B^\mu)$, 
would generate an effective chemical potential for lepton (or baryon) number~\cite{Cohen:1987vi,Cohen:1988kt, Dolgov:1994zq,Dolgov:1996qq}. Therefore once the lepton (or baryon) number  is generated by certain processes in the early universe, the chemical potential will contribute to the generation of net lepton (or baryon) number.  We note that similar derivative couplings (or variations) have been adopted to explain baryon asymmetry in earlier works~\cite{Takahashi:2015ula,Domcke:2020kcp,Barrie:2020hiu}.

In this work, we consider Higgs inflation as a concrete example~\cite{Salopek:1988qh,Bezrukov:2007ep}. Two higher dimensional operators in the framework of Higgs inflation: First, we consider the dimension-five Weinberg operator, violating lepton number~\cite{PhysRevLett.43.1566}: ${\cal O}_{\rm dim 5} \sim (L\phi)(L\phi)^\dagger$.\footnote{One of the ultraviolet (UV) completion models is the heavy Majorana neutrino being integrated out, motivated by the seesaw mechanism.} Second,  the derivative coupling between the Higgs field $ \phi $ and the lepton number current $ j_{L}^{\mu} $: ${\cal O}_{\rm dim 6} \sim  \partial_\mu (\phi^\dagger \phi) j_L^\mu$~\cite{Cohen:1987vi,Cohen:1988kt, Dolgov:1994zq,Dolgov:1996qq}. As long as we accept the coherence of the Higgs field for spatial variation (i.e. $\partial_i \phi =0$), the dimension six operator generates effective (time dependent) chemical potential $\mu_{\rm eff}(t) \sim \partial_t (\phi^\dagger \phi)$ as ${\cal O}_{\rm dim 6} \sim \partial_t (\phi^\dagger \phi) (n_\ell - n_{\overline{\ell}})$ for the difference in lepton and anti-lepton number densities.  In our setup, all the masses of the SM particles including neutrinos change in the course of time by oscillating Higgs field after the inflation, which allows generation of lepton number as we will closely see in the later sections.

This paper is organized as follows: In Section \ref{Setup}, we will set out model including the higher dimensional operators. In Section \ref{Neutrino Production}, we calculate particle production and subsequent lepton asymmetry from the coherently oscillating Higgs inflaton background. We also present the numerical results and read the viable parameter space to explain current baryon asymmetry. In Section \ref{Conclusion and Discussion}, we conclude.

\section{Setup} \label{Setup}
In this section, we will briefly review the major results of the Higgs inflation~\cite{Salopek:1988qh,Bezrukov:2007ep}, as well as relevant characteristics of reheating phase after inflation~\cite{Kofman:1997yn,GarciaBellido:2008ab,Bezrukov:2008ut}. Then we will discuss additional two higher dimensional operators to the SM Lagrangian, and their effects to baryon asymmetry.

\subsection{Higgs Field Equation during and after Inflation}
Higgs inflation has attracted a lot of attention due to its simplicity and consistency to the current measurement of the spectral index $ n_{s} $ and tensor-to-scalar ratio $ r $. To suppress $ r $ value, one introduces a non-minimal coupling between the Higgs scalar and the Ricci scalar $ R $. One of the distinguished features of the Higgs inflation is that the fermion masses are very large and time-dependent during the inflation and subsequent reheating epochs due to the dynamics of the vacuum expectation value of the Higgs field as an inflaton. We emphasize that this is also responsible for the large lepton asymmetry generated from the Higgs inflaton condensate.\footnote{See Ref.~\cite{Pearce:2015nga} where lower reheating temperature of the order of $ 10^{12} ~\rm GeV$ is assumed in order to generate sufficient lepton number to photon ratio in a non-Higgs inflation. In Ref. \cite{Enomoto:2020lpf} oscillating Higgs field during (p)reheating is considered with preexisting C/CP violation as the source of asymmetry.}

In addition, to describe the action in the canonical Einstein frame without the non-minimal coupling, one performs the conformal transformation. From this, fermion masses also obtain additional conformal weight. See Appendix~\ref{App:ConfTrsf} for detailed derivation.

The Jordan-frame action of the inflaton sector is given by the SM Higgs with a non-minimal coupling constant $ \xi $:
\begin{align} \label{eq2.1}
	S_{J,\rm inf} = \int d^{4}x \sqrt{-g_{J}} \left[\frac{1}{2}\left( M_{P}^{2} + \xi \phi_{J}^{\dagger} \phi_{J} \right) R_{J} 	- \frac{1}{2}\vert  \partial_{\mu} \phi_{J} \vert^{2} - V_{J}(\phi_{J})	\right]
\end{align}
where $ J $ stands for the Jordan frame, $ V_{J}(\phi_{J}) = \frac{\lambda}{4} \phi_{J}^{4} $, and the real scalar $\phi_J$ is the physical Higgs direction in the unitary gauge. We transform the action Eq.~(\ref{eq2.1}) to Einstein frame with Weyl transformation:
\begin{align}
	g_{\mu\nu} =\Omega^{2}(\phi_{J}) g_{J\mu\nu},
\end{align}
where we truncate the conformal factor $ \Omega^{2} $ up to the quadratic power of $\phi_J$:
\begin{align} \label{Eq:Conformal}
	\Omega^{2}(\phi_{J})\equiv 1 + \frac{\xi}{M_{P}^{2}} \phi_{J}^{2}.
\end{align}
We have omitted explicit $ E $ subscript for simplicity.
The final form of the action in the Einstein frame is
\begin{align}
	S_{E,\inf} = \int d^{4}x \sqrt{-g} \left[ \frac{M_{P}^{2}}{2} R - \frac{1}{2} \vert \partial_{\mu} \phi \vert^{2} - V (\phi)\right],
\end{align}
where the potential in the Einstein frame $ V(\phi) $ is
\begin{align}
	V(\phi) = \frac{V_{J}(\phi_{J}(\phi))}{\Omega^{4}}.
\end{align}
The inflaton fields in both frames are related by
\begin{align} \label{Eq:JtoE}
	\frac{\phi}{M_{P}} = \sqrt{\frac{1+6\xi}{\xi}} \sinh^{-1} \left(	\frac{\sqrt{\xi(1+ 6 \xi)}\phi_{J}}{M_{P}}	\right) - \sqrt{6} \sinh^{-1} \left(	\sqrt{\frac{6\xi^{2} \phi_{J}^{2} / M_{P}^{2}}{1 + \xi \phi_{J}^{2} / M_{P}^{2}}}	\right).
\end{align}
In the limit of $ \xi \phi \gg M_{P} $, Eq.~(\ref{Eq:Conformal}) and Eq.~(\ref{Eq:JtoE}) can be approximated as
\begin{align}
\Omega^{2}\simeq
e^{\sqrt{\frac{2}{3}} \frac{\vert \phi \vert}{M_{p}}},
&\frac{\phi}{M_{P}}
	&\simeq \left(\sgn\phi_J\right)
\frac{\sqrt{6}}{2} \ln \left(	1 + \xi \frac{\phi_{J}^{2}}{M_{P}^{2}}	\right).
\end{align} 
Therefore,
\begin{align}
	\frac{\phi_{J}}{M_{P}}
		\simeq
		\left(\sgn\phi\right)\frac{1}{\sqrt{\xi}} \left(	e^{\sqrt{\frac{2}{3}} \frac{\vert \phi \vert}{M_{P}}} -1	\right)^{1/2}.
\end{align}

Let $\phi_0$ be the inflaton field value at the end of inflation, after which the inflaton potential can be approximated by a quadratic form in general:\footnote{This approximation to quadratic potential form is generically valid  in Higgs inflation when $ \frac{M_{P}}{\xi} \ll  \phi \lesssim \frac{M_{P}}{\sqrt{\xi}} $, corresponding to early stages of the reheating where the amplitude of oscillation is in intermediate range.  The value of $\phi$ stays large ($\phi \gg M_P/\xi$) most of the time, and passes by smaller field value regime only in short periods of time. After several oscillations the amplitude becomes small $\phi \lsim M_P/\xi$ and the potential is reduced to the conventional quartic one in lower energies. More details are found in Appendix \ref{App:ConfTrsf}.}
\begin{align} \label{Eq:quadraticpotential}
	V(\phi) \simeq	\frac{1}{2}M^{2}\phi^{2}.
\end{align}
Here, $ M $ determines the frequency of the oscillation right after the inflation. Then the solution of the Friedmann equations
\begin{align}
	\frac{1}{2} \dot{\phi}^{2} + V(\phi) &= 3 M_{P}^{2} H^{2} \\
	\ddot{\phi} + 3 H \dot{\phi} + \frac{d V}{d\phi} &= 0 
\end{align}
describes the dynamics of the oscillating homogeneous inflaton field as an approximate form~\cite{Kofman:1997yn}
\begin{align} \label{Eq:AppSol}
	\phi(t) \approx \phi_{0} \frac{\sin (Mt)}{Mt}.
\end{align}

In the Higgs inflation, the effective mass $ M $ during the oscillation is not a free parameter but fixed from $ \lambda $ and $ \xi $ as $ M^{2} = \frac{\lambda M_{P}^{2}}{3 \xi^{2}} $, and $\phi_0$ is order of Planck scale~\cite{GarciaBellido:2008ab}. As we will see, the neutrino mass and the effective chemical potential of lepton number change in time accordingly to the oscillating Higgs field in Eq.~(\ref{Eq:AppSol}).

\subsection{Lepton Number Violation and Effective Chemical Potential}

In the SM, it is the simplest to introduce the Weinberg operator, the minimal lepton-number-violating dimension-five operator, in order to account for the observed neutrino masses:
\begin{align} \label{Eq:Weinberg}
\mathcal{L}_{\text{dim-5},J} = \frac{c_{5}}{M_{P}} (\overline{L}_{J} \widetilde{\Phi}_{J})(\widetilde{\Phi}_{J} L_{J})^{\dagger},
\end{align}
where $ \widetilde{\Phi} \equiv i \sigma_{2} \Phi^{*} $ and we have assumed this operator is Planck suppressed with a parameter $ c_{5} \lesssim \mathcal{ O}(1) $.
This can be used to realize a leptogenesis.
We emphasize that we are in the Jordan frame now. Although we will only consider a single flavor in the following discussion, the generalizations to three flavor or more are straightforward. 

This operator generates the neutrino mass 
\begin{align}
m_{\nu} = \frac{c_{5}}{M_{P}} \frac{\phi_{J}^{2}}{\Omega} = \frac{2 c_{5} M_{P}}{\xi}\sinh \left(\frac{\vert \phi \vert}{\sqrt{6}M_{P}}\right)
\label{Eq:NeutrinoMassinEframe}
\end{align}
in Einstein frame, where we take conformal transformation of the fermion mass $ m \rightarrow m/\Omega $ into account. The detailed derivation of Eq.~(\ref{Eq:NeutrinoMassinEframe}) is found in Appendix \ref{App:ConfTrsf}.

On the other hand, during and right after the inflation, the fact that the inflaton field is nearly homogeneous implies $ \partial_{\mu} \phi  = \dot{\phi} \delta_\mu^0$. As mentioned in Introduction, this inflaton background  spontaneously breaks Lorentz symmetry, and hence CPT symmetry. We emphasize that this is universal characteristic of all standard inflation scenarios and motivates us to think the spontaneous symmetry breaking of CPT as the source of matter-antimatter asymmetry in the inflationary framework.

As we will see, when inflaton couples to derivative of a lepton (or baryon) current $ j_{L}^{\mu} = \overline{\ell} \sigma^{\mu} \ell$, the CPT breaking actually leads to asymmetry between decay rates into particles and antiparticles. In this work, we introduce an effective dimension-six operator~\cite{Pearce:2015nga}:
	\begin{align}
	\mathcal{ O}_{6}
	= - \frac{c_{6}}{M_{P}^{2}} \Phi_{J}^{\dagger} \Phi_{J} \partial_{\mu} j^{\mu}_{L} 
	= \frac{c_{6}}{M_{P}^{2}}  (\partial_{\mu}\phi_{J}^{2} ) j_{L}^{\mu} 
	= \frac{c_{6}}{M_{P}^{2}} (\partial_{t} \phi_{J}^{2}) j^{0}_{L},
	\label{Eq:chemical potential operator}
	\end{align}
where we have taken unitary gauge to the Higgs doublet $ \Phi $ and the dimensionless constant $ c_{6} $ parameterizes the size of the dimension-six effective coupling.
In the last equation, we used the spatial homogeneity of the inflaton sector. $ j_{L}^{0} $ has the meaning of the number density of the lepton, $ j_{L}^{0} \equiv n_{L} = l^{\dagger} l = n_{\ell} - n_{\overline{\ell}}$. Note that the operator Eq.~(\ref{Eq:chemical potential operator}) does not violate the lepton number by itself, though it generate a difference of decay rates between particles and anti-particles under presence of another lepton-number violating operator. Therefore, we introduce explicit lepton number violating Weinberg operator as discussed earlier.\footnote{However, there is another source of lepton number violation in Eq.~(\ref{Eq:chemical potential operator}): anomalous $ B+L $. In general, the global $L$ is anomalous and its current $j^\mu_L$ is not conserved.
Indeed in the SM, what is anomalous is $B+L$, and therefore the anomaly-induced higher-dimensional operator from $\Phi^\dagger\Phi F\widetilde{F}$ should rather be $\Phi_J^\dagger\Phi_J\partial_\mu(j_L^\mu+j_B^\mu)$. Without additional source of $ B-L $ breaking, the generated $ B+L $ by this operator would be quickly washed out by the sphaleron effects in the absence of strong first order phase transition.}

\begin{figure}[t]\center
	\includegraphics[width=7.5cm]{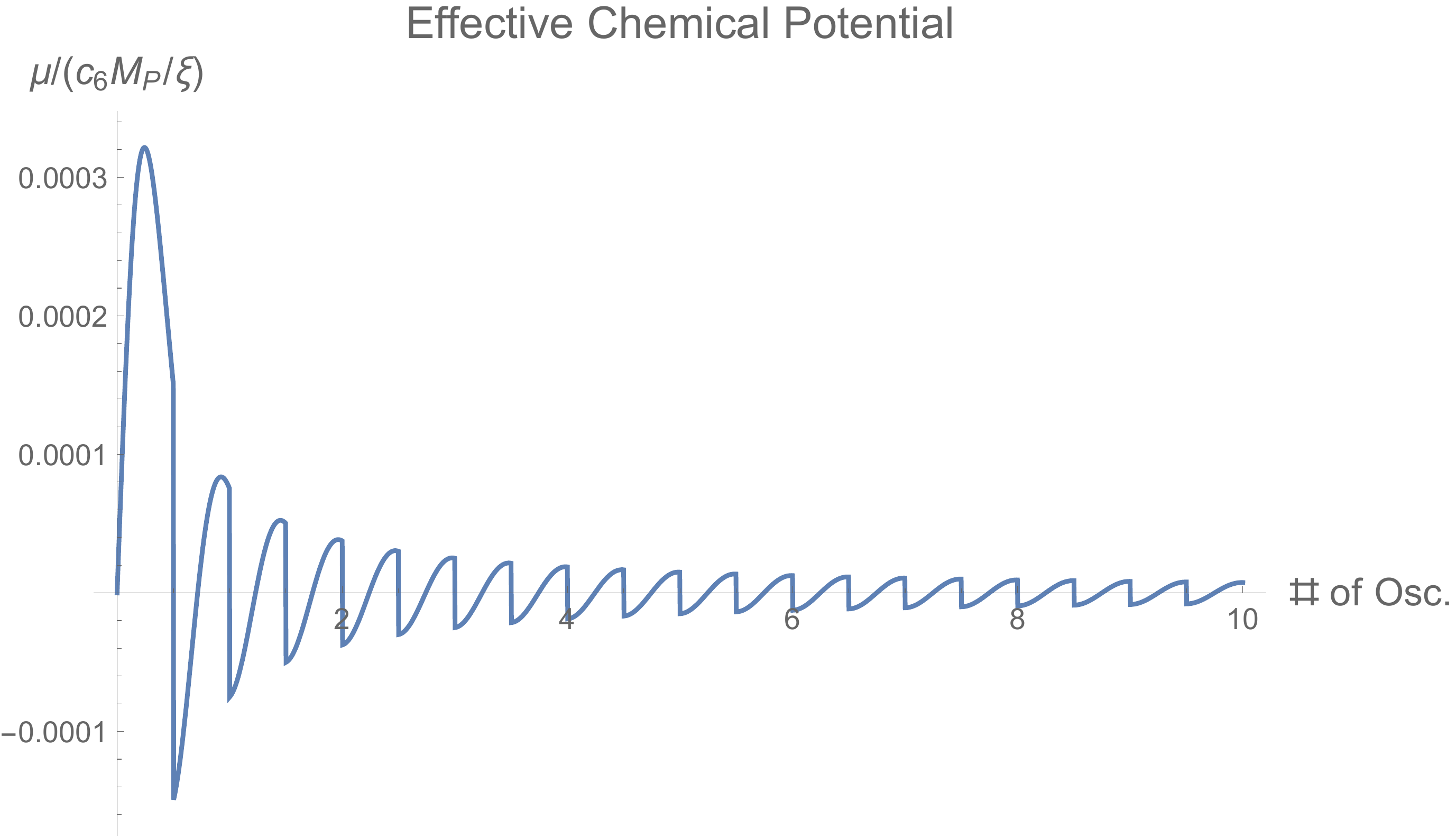}
	\includegraphics[width=7.5cm]{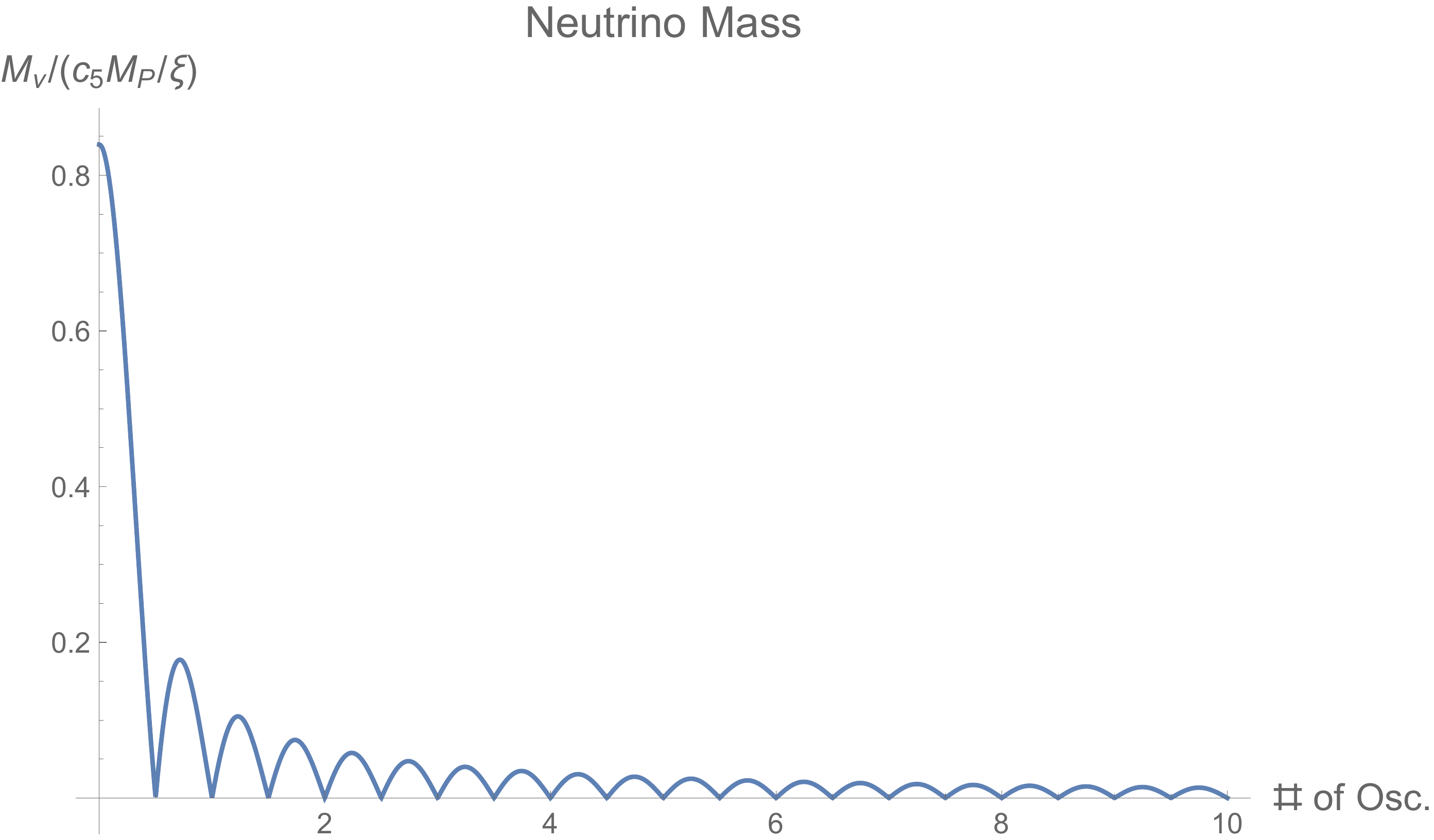}
	\caption{The shape of the effective chemical potential (left) and the neutrino mass (right) as functions of number of oscillation: $ M t/ 2\pi $, where $ t $ is the physical time in Einstein frame. The values are rescaled by relevant factors $ c_{5}M_{P} / \xi $ and $ c_{6}M_{P} / \xi $, respectively. Both are sizable only during the first few oscillations.
	} \label{Fig:eff}
\end{figure}

A possible intuitive interpretation of the operator~Eq.~\eqref{Eq:chemical potential operator} is as an effective chemical potential in the sense that this term makes energy density shift, $ \Delta \rho = - \frac{c_{6}}{M_{P}^{2}} (\partial_{t} \phi^{2}) j_{L}^{0} $ \cite{Cohen:1987vi,Cohen:1988kt,DeSimone:2016ofp}, although this picture is under controversy; see Refs.~\cite{Dolgov:1996qq,Dasgupta:2018eha}. Note that our result of asymmetry between particle and anti-particle is valid no matter whether this interpretation is possible or not.
Anyway motivated by this interpretation, we call the following the ``(effective) chemical potential'':
\begin{align}
	\mu_{\rm eff}(\phi_{J}) \equiv  \frac{c_{6}}{M_{P}^{2}} (\partial_{t} \phi_{J}^{2}).
\end{align}
As we will see, this makes difference between dispersion relations of particle and antipaticle. Due to the exact cancellation of the conformal factor from the metric and current, the effective chemical potential does not obtain additional conformal factor. However, we still need to consider the inflaton field transformation
\begin{align} \label{Eq:Chemical}
\mu_{\rm eff}(\phi) = \frac{c_{6}}{\xi}  \partial_{t}  \left(  e^{\sqrt{\frac{2}{3}} \frac{\vert \phi \vert}{M_{P}}}  \right) 	.
\end{align}
Fig.~\ref{Fig:eff} shows shape of the chemical potential and the neutrino mass as a function of the number of oscillations.

\section{Neutrino Production from Oscillating Higgs Inflaton} \label{Neutrino Production}
Due to its coherence of the field configuration of the inflaton field after the inflation, it is legitimate to describe the inflaton field as a classical field, but the matter field should be quantized. Therefore, we are describing Higgs condensate as a time dependent classical field and only quantize neutrino fields. 

\subsection{Lepton Number Density}

In the oscillating phase, the Higgs field varies in time and results in oscillations of SM particle masses, especially of neutrino ones that are induced by lepton number violating Weinberg operator in Eq.~(\ref{Eq:NeutrinoMassinEframe}). The generated numbers of neutrinos and anti-neutrinos can be obtained by a method analogous to the gravitational particle production in time dependent background~\cite{Birrell:1982ix,Mukhanov:2007zz} following Ref.~\cite{Pearce:2015nga}. The details of calculations are found in Appendix~\ref{App:BogoTrsf}, and we only briefly sketch the ideas and some key results here.

The equation of motion  for neutrino with the chemical potential term in Eq.~(\ref{Eq:chemical potential operator}) in momentum space is: 
\begin{align}
	( i \partial_{\tau} + \vec{\sigma} \cdot \vec{k} ) \nu_{L} = - \widetilde{m}_{\nu} (i \sigma_{2}) \nu_{L}^{*} - \widetilde{\mu} \nu_{L}.
	\label{eq:eom}
\end{align}
Here we use the conformal time defined  as $ \tau \equiv \int dt /a(t)$ and the comoving momentum $ k \equiv a p $, and mass parameters $ \widetilde{m}_{\nu} \equiv a m_{\nu}$ and $\widetilde{\mu} \equiv a \mu $ with $ a $ and $p$ being the scale factor and the physical momentum, respectively.  We also define time derivatives as  $ F^{\prime} \equiv \partial_{\tau} F $ and $ \dot{F} \equiv \partial_{t} F$ for any function $ F $.

\begin{figure} \centering
	\includegraphics[width=13cm]{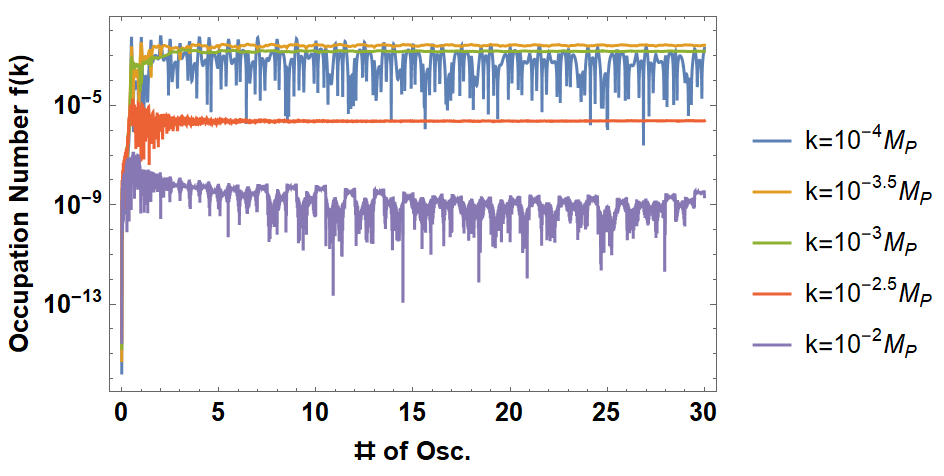}
	\caption{The time evolution of the absolute values of occupation number $f(k)$ of the neutrino for various $k$ values with $\xi=100$ and $c_{5} = c_{6} = 1$. For large $k$, the occuption number converges rapidly and is highly suppressed, so the contribution to the asymmetry is small. The asymmetry is mainly generated at the early stages of the reheating within a few oscillation $ \mathcal{O}(1) $ and converges soon after the end of the inflation. For smaller $ k $, even though the convergence is rather weak, the total contribution is minor due to small phase space.} \label{Fig:number}
\end{figure}

\begin{figure}\centering
	\includegraphics[width=15cm]{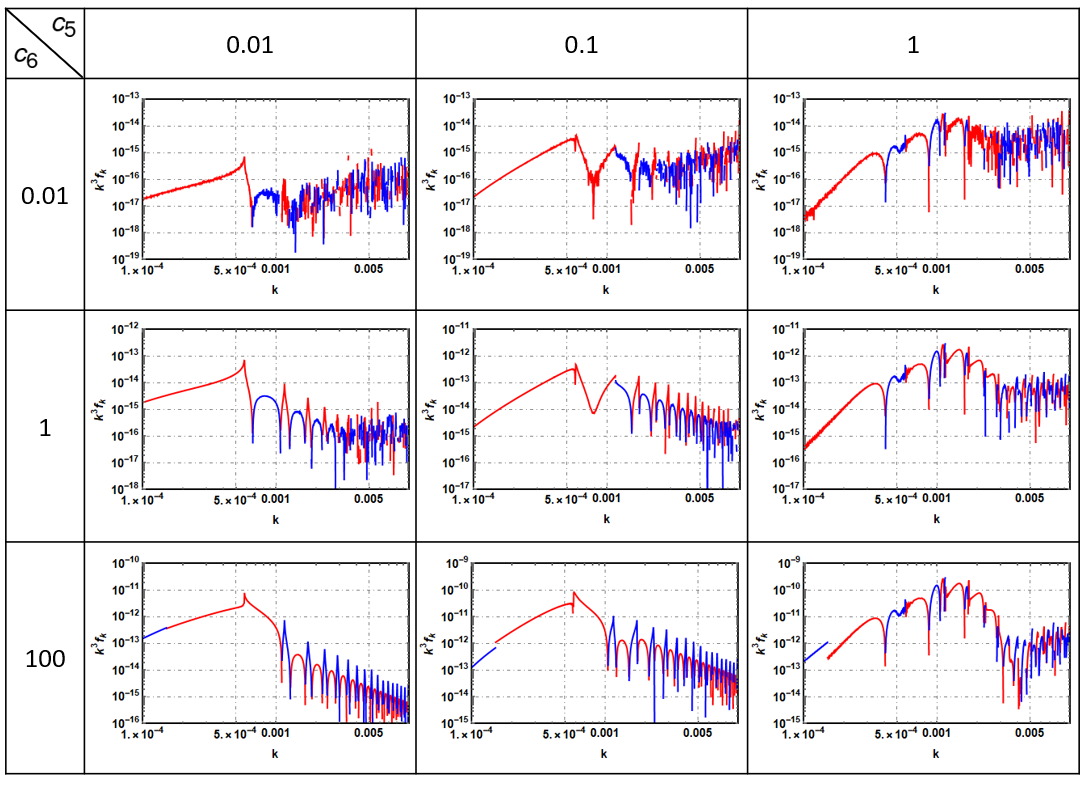}
	\caption{The final occupation numbers multiplied by $k^{3}$ considering the contribution of from the phase space depending on the $ (c_{5},c_{6}) $. Blue and red colors are used to represent different sign of the final occupation numbers.} \label{Fig:spectrum}
\end{figure}

By  Bogoliubov transformation, Eq.~(\ref{eq:eom}) leads to a coupled equations of Bogoliubov coefficients $ \alpha $ and $ \beta $,
\begin{align} \label{Eq:coupled}
	\alpha_{s}^{\prime}(\tau,k) = - \frac{\beta_{s}(\tau,k)}{2\omega_{s}^{2}} \left[	\widetilde{m}_{\nu} \widetilde{\mu}^{\prime} - (s k + \widetilde{\mu}) \widetilde{m}_{\nu}^{\prime}	\right] e^{2 i \int^{\tau}_{0} \omega_{s}(\tau^{\prime}) d\tau^{\prime}}, \nonumber\\
	\beta_{s}^{\prime}(\tau,k) =  \frac{\alpha_{s}(\tau,k)}{2\omega_{s}^{2}} \left[	\widetilde{m}_{\nu} \widetilde{\mu}^{\prime} - (s k + \widetilde{\mu}) \widetilde{m}_{\nu}^{\prime}	\right] e^{-2 i \int^{\tau}_{0} \omega_{s}(\tau^{\prime}) d\tau^{\prime} },
\end{align}
with the initial conditions $ \alpha_{s}(0,k) = 1 $ and $ \beta_{s}(0,k)=0 $, where $ s = \pm $ represents helicity states. These equations are numerically solved from $ \tau=0 $ at the end of the inflation. Importantly,  non-zero $ \beta_{s} $ values at late times indicate particle production from the time dependent background: the generated number density for the helicity $ s $ is given by
\begin{align} 
n_{s}(t) = \frac{1}{(a(t)/a_{\rm end})^{3}} \int \frac{d^{3} k}{(2\pi)^{3}} \vert \beta_{s}(\tau(t),k) \vert^{2},
\end{align}
where  $ \beta_{s}(\tau,k) $ is Bogoliubov coefficient from the time-dependent background or the change of vacuum state at each moment obtained by Eq.~(\ref{Eq:coupled}) and $ a_{\rm end} $ is the scale factor at the end of the inflation. Note that $ a(t)^{-3} $ factor in front accounts dilution from the expansion of the universe.
The occupation number for $s$ at the end of reheating is $ f_{s}(t,k)= \vert \beta_{s}(\tau(t),k) \vert^{2}$. The final net number density of the produced lepton is the difference between the number densities of each helicity states at a sufficiently late time when the Higgs field value becomes small so that $ m_{\nu}(\phi) \ll T_{\rm reh} $ \cite{Pearce:2015nga}:
\begin{align}
n_{L} \vert_{\rm reh} \equiv \lim_{t \rightarrow t_{\rm reh}} n_{\ell}(t)-n_{\overline{\ell}}(t)= 
\left(\frac{a_{\rm end}}{a_{\rm reh}}\right)^{3} 
\int \frac{d^{3}k}{(2\pi)^{3}} f(k),
\end{align}
where the net occupation number is the helicity weighted sum at the end of the reheating $ f(p) = f(t_{\rm reh},k) = \sum_{s} s f_{s}(k) = f_{+} - f_{-}$ describing the differences between particle ($s=+$) and anti-particle ($s=-$). Comoving lepton number density $  \widetilde{n}_{L} \equiv \int \frac{d^{3}k}{(2\pi)^{3}} f(k) $ is insensitive to $ t_{\rm reh} $ as $ f(t,k) $ rapidly approches to a constant for a given $ k $. The explicit evolution of the occupation numbers $ f(t,k) $ for various modes are depicted in Fig.~\ref{Fig:number}. Note that the asymmetry is dominantly generated during the early time within first several oscillations so that it is relatively insensitive to exact value of $ t_{\rm reh} $.  $ \widetilde{n}_{L} $ is determined by $ c_{5} $ and $ c_{6} $ paramters. Final spectrum of the occupation number $ f(k) $ multiplied by $ k^{3} $ taking the contribution from the phase space are also shown in Fig.~\ref{Fig:spectrum}. For large $ k $ modes, the occupation numbers are suppressed. For small modes, even though the occupation numbers are larger, the phase space is too small to contribute to the final asymmetry.

\subsection{Lepton Asymmetry and Reheating Temperature}

\begin{figure}[t]
	\centering
	\includegraphics[width=12cm]{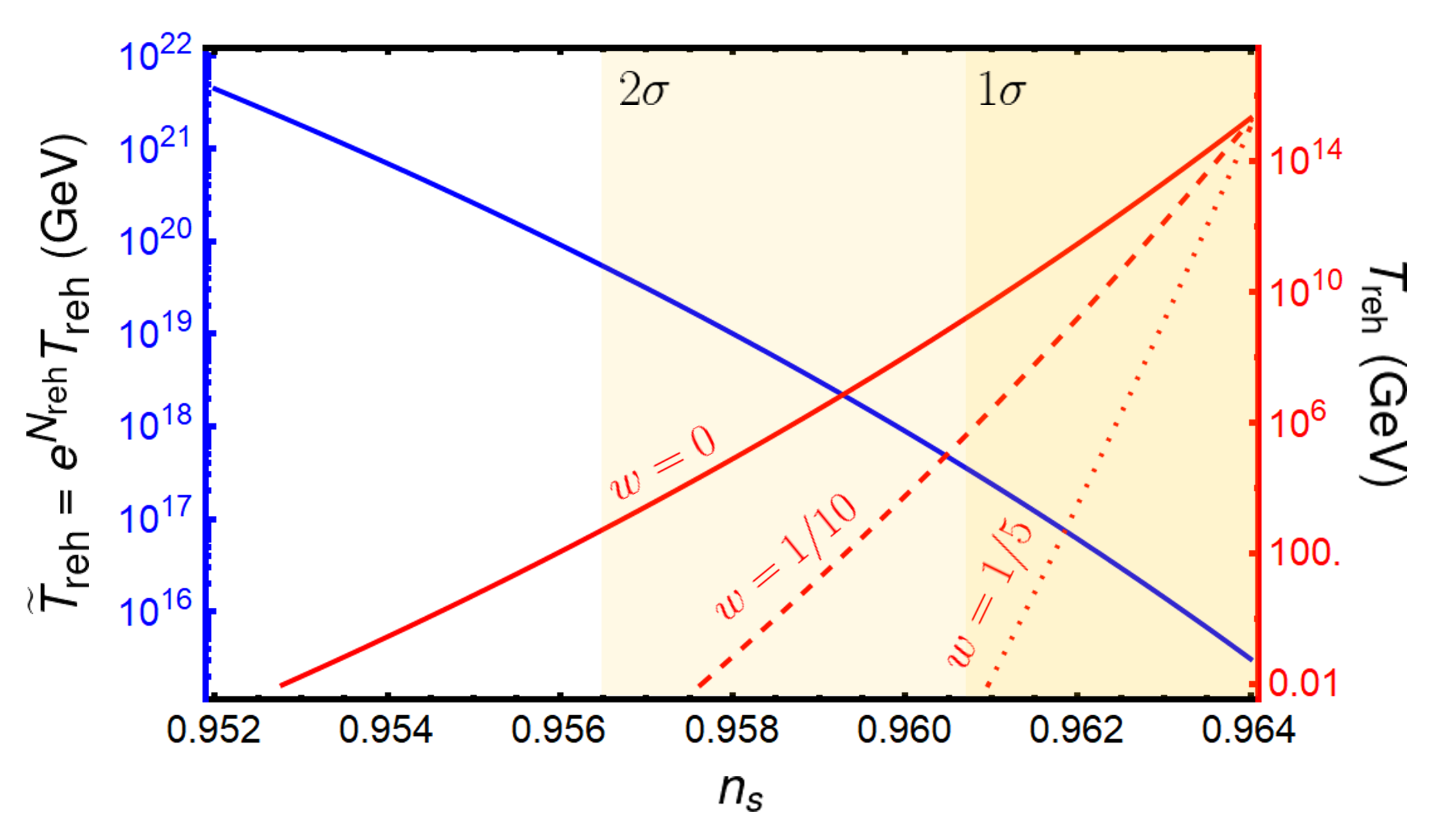}
	\caption{Rescaled temperature $\widetilde{T} \equiv e^{N_{\rm reh}} T_{\rm reh}$ and $T_{\rm reh}$ vs $n_s$. The blue line is $\widetilde{T} $ and red solid, dashed, and dotted lines represent the reheating temperatures $ T_{\rm reh} $ assuming $ w = (0,1/10,1/5) $ during the reheating, respectively. Results are calculated taking Ref.~\cite{Cook:2015vqa} into account by combining $ N_{\rm reh}$ and $T_{\rm reh} $ in the case of the Higgs inflation. Unlike the reheating temperature or e-folding number of the reheating (red lines), blue line is independent of the choice of the equation of state. Orange colored regions are $1\sigma$ and $2\sigma $ allowed regions from the Planck 2018 result on the spectral index $ n_{s} $ \cite{Akrami:2018odb}.} \label{Fig:aTbound}
\end{figure}

Finally, the lepton asymmetry is given by
\begin{align} 
\eta_{L}(t_{\rm reh}) &\equiv \left. \frac{n_{L}}{n_{\gamma}} \right\vert_{\rm reh} = \frac{\pi^{2}}{2 \zeta(3)} \left.\frac{\widetilde{n}_{L}}{\widetilde{T}^{3}} \right\vert_{\rm reh}
\end{align}
where the photon number density at the end of the reheating $ {n_{\gamma}\vert_{\rm reh} = \frac{2\zeta(3)}{\pi^{2}} T_{\rm reh}^{3} } $ with $ \zeta(3) \simeq 1.2 $ and $ \widetilde{T} \equiv e^{N_{\rm reh}} T_{\rm reh} $ where $N_{\rm reh} = \ln \left( \frac{a_{\rm reh}}{a_{\rm end}} \right) $.\footnote{Because the interaction rate due to the dimension-five operator \ref{Eq:Weinberg} is small compared to the expansion rate, there will be no further lepton number production after the reheating. When we regard Majorana neutrino as the origin of dimension-five operator, this is equivalent to impose $ T_{\rm reh} < M_{N} $, which is  satisfied in Higgs inflation when $ M_{N} \gtrsim \mathcal{O}(10^{15}~\rm GeV) $. See Figure \ref{Fig:aTbound}.}
Because $ \widetilde{n}_{L} $ is solely determined by $ c_{5} $ and $ c_{6} $, it suffices to specify $ \widetilde{T} $ (which we call `rescaled temperature') to obtain $ \eta_{L}(t_{\rm reh}) $. Although (p)reheating processes after the end of inflation is highly non-trivial and model dependent,\footnote{Early analysis relevant for the reheating in the Higgs inflation is done in Refs. \cite{Bezrukov:2008ut,GarciaBellido:2008ab,Repond:2016sol, DeCross:2015uza}. For the possible unitarity violation issue during the Higgs inflation reheating, see Refs. \cite{Ema:2016dny,Hamada:2020kuy}.
In particular, Ref. \cite{Ema:2016dny} argued that the longitudinal mode of gauge boson production is so violent that preheating may finish right after the first oscillation in Higgs inflation with a large non-minimal coupling $ \xi \gtrsim 100 $. However, this phenomena violates the unitarity but also is sensitive to higher order operators \cite{Hamada:2020kuy}. Therefore, we will assume there are at least a few oscillations after the inflation in our work. } we can find possible parameter regions of the rescaled temperature without large uncertainties \cite{Cook:2015vqa} as we will see below. See Appendix \ref{App:RelationTemp} for more detailed review.

Rescaled temperature $ \widetilde{T} \equiv e^{N_{\rm reh}}T_{\rm reh} $ can be inferred by the Hubble scale $ H_{*} $ at pivot scale and e-folding number $ N_{*} $ between time when the pivot scale $ k_{*} $ crosses the horizon and the time at the end of the inflation without the knowledge of the equation of state during the reheating:
\begin{align}
\widetilde{T} = e^{N_{\rm reh}}T_{\rm reh} = \left(	\frac{43}{11 g_{*,\rm reh}}	\right)^{\frac{1}{3}}
\left(\frac{a_{0} T_{0}}{k_{*}}\right) H_{*} e^{-N_{*}}.
\end{align}
For a given model of inflation, $ H_{*} $ and $ N_{*} $ are generally related to the slow-roll parameters $ \epsilon $ and $ \eta $, hence the observables $ n_{s} $ and $ r $ at pivot scale. In the case of Higgs inflation \cite{Bezrukov:2007ep},
\begin{align}
\epsilon_{*} \simeq \frac{3}{4 N_{*}^{2}},&&\eta_{*} \simeq -\frac{1}{N_{*}},&& n_{s} \simeq 1 - 6 \epsilon_{*} + 2 \eta_{*} \simeq 1- \frac{2}{N_{*}},&&N_{*} \simeq \frac{2}{1-n_{s}}. 
\end{align}
The Hubble scale also determines the amplitude of the power spectrum is given by
\begin{align}
A_{s} = \frac{1}{8 \pi^{2} \epsilon_{*}} \frac{H_{*}^{2}}{M_{P}^{2}},&&H_{*} = \pi M_{P} \sqrt{\frac{3}{2}A_{s}} (1-n_{s}).
\end{align}
Hence, we can evaluate $ \widetilde{T}$ solely as a function of $ n_{s} $.
Fig.~\ref{Fig:aTbound} depicts the values of $ \widetilde{T} $ and $ T_{\rm reh} $ as a function of the spectral index $ n_{s} $. Note that rescaled temperature $ \widetilde{T} $ is independent of equation of state, $w=p/\rho$ during the reheating while $ T_{\rm reh} $ itself is highly depends on $ w $~\cite{Gong:2015qha,Cai:2015soa,Cook:2015vqa}. From the Planck 2018 1$ \sigma $ bound, we will consider $\widetilde{T} \in (10^{15} - 10^{18}) {\rm GeV}$ as a preferred regime. Assuming $ w = 0 $, the reheating temperature suggested in Ref.~\cite{Bezrukov:2008ut} $ T_{\rm reh} \in (3\times 10^{13}-1.5 \times 10^{14})~\rm GeV$ corresponds to $ \widetilde{T} \simeq 10^{16} ~\rm GeV $, which is also within the preferred regime.

\subsection{Baryon Asymmetry}
In our scenario, `net' $ B-L $ number (in fact, only $ L $ number) is generated right after the inflation and is frozen by the time of the end of reheating. Later, generated lepton number asymmetry transfers to baryon number through sphaleron effects, conserving $ B-L $, so that
\begin{align}
n_{B}
&= C_{\rm sphal}\, n_{B-L,i},	&
C^{\rm (SM)}_{\rm sphal}
&= \frac{28}{79}
\end{align}
\cite{Khlebnikov:1988sr}. Therefore, for successful baryon asymmetry, we also require $ \eta_{L,i} \gtrsim 3 \eta_{B}$.

\begin{figure}
	\includegraphics[width=15cm]{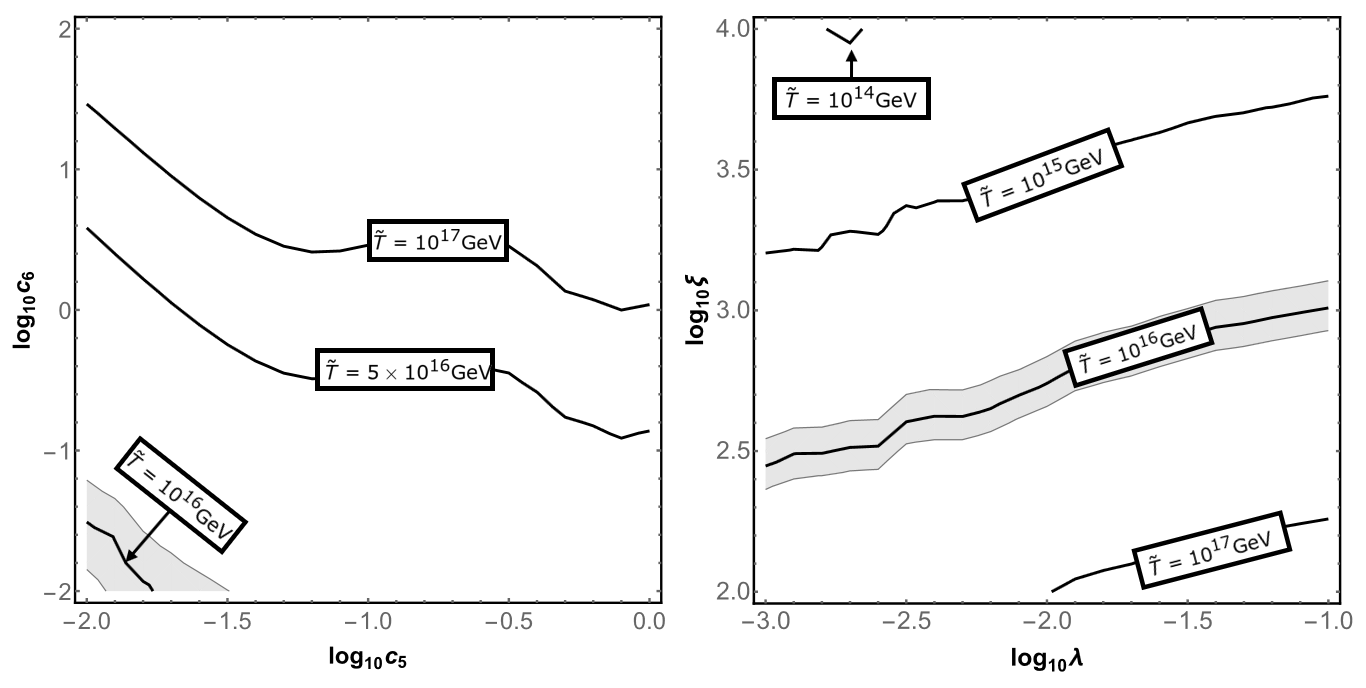}
	\caption{
	Solid lines represent the parameter values of $ (c_{5}, c_{6}) $ (left)
	$ (\lambda,\xi) $ (right) explaining current baryon asymmetry $ \eta_{B} \simeq 6 \times 10^{-10} $ for various rescaled temperatures denoted in the box. We have taken $ (c_{5},c_{6}) = (1,1) $ for the left panel and $ (\lambda,\xi) = (10^{-2},100) $ for the right panel. Shaded regions correspond to the reheating temperature given in Ref.~\cite{Bezrukov:2008ut} with $ w=0 $.}  \label{Fig:result}
\end{figure}

Considering the transition through the sphaleron effect, final baryon asymmetry for given comoving lepton number $ \widetilde{n}_{L}(c_{5},c_{6}) $ and rescaled reheating temperature $ \widetilde{T} $ is given by
\begin{align} \label{Eq:Baryonasymm}
\eta_{B} (c_{5},c_{6},\widetilde{T}) \simeq \frac{C^{\rm (SM)}_{\rm sphal} \pi^{2} \widetilde{n}_{L}}{2 \zeta(3) \widetilde{T}^{3}}.
\end{align}
The observed value is $ \eta_{B,\rm obs} \simeq 6 \times 10^{-10} $ \cite{Zyla:2020zbs}. As can be seen from Fig.~\ref{Fig:result}, there exist reasonable parameter regions of ($ c_{5},c_{6} $) and $ (\lambda,\xi) $ in the Higgs inflation scenario which explain current baryon asymmetry $ \eta_{B} $ with desired rescaled temperature.

To avoid unitarity violation from larger $ m_{\nu} $ or $ \mu_{\rm eff}  $ than cut-off scale of Higgs inflation during the reheating $\Lambda = M_{P}/\xi $ \cite{Burgess:2009ea,Barbon:2009ya,Burgess:2010zq,Lerner:2009na,Bezrukov:2010jz}, we will only consider the cases where $ c_{5} \lesssim 1 $ and $ c_{6} \lesssim 100 $. The particles with the momentum larger than the cut-off would be strongly coupled and their behavior cannot be descibed within the effective field theory. In our parameter choice, the majority of the particles are generated with momentum smaller than the cut-off scale $\Lambda = M_{P}/\xi $ as depicted in Fig.~\ref{Fig:spectrum}.\footnote{
In Fig.~\ref{Fig:spectrum}, the Einstein-frame momentum cut-off is $\Lambda= M_{P} / \xi =0.01 $. Large $ c_{6} \sim 100$ makes it one order of magnitude smaller, $ \Lambda/\sqrt{c_{6}}\sim 10^{-3}$. We see that the unitarity issue arises, namely the location of dominant peaks becomes located at $p\gtrsim 10^{-3}$, only when $ c_{5} \gtrsim 1 $ and $ c_{6} \gtrsim 100 $.
}
For larger $ \xi \gtrsim 10^{4} $, the suppression on lepton number density so large that the asymmetry may not be enough to explain current relic within preferred temperature regime. This can be seen in Fig.~\ref{Fig:result}.

Usually, it is believed that the spontaneous baryogenesis mechanism does not work in super-cooled regime due to its inefficiency from the cancellation during the oscillations \cite{DeSimone:2016ofp}. However, as identifying the inflaton decay as the source of the lepton asymmetry, both the amplitude of the effective chemical potential and the neutrino mass are so large that the produced lepton numbers are enough to explain current relic density without any fine-tuning of the parameters. We also do not have to assume such a low scale inflation as in Ref.~\cite{Pearce:2015nga}, which considers neutrino decay from the condensate of the SM Higgs (not identified as the inflaton).

As the final remark, we note that the strong bound on spontaneous baryogenesis from baryon isocurvature perturbation~\cite{DeSimone:2016ofp} is not imposed in our scenario because we do not assume any light bosonic fields in the early time other than SM. This is one of the advantages regarding the possibility of the inflaton decay as the source of baryon asymmetry. 

\section{Conclusion and Discussion} \label{Conclusion and Discussion}
We propose a scenario explaining baryon asymmetry of our universe in Higgs inflation.  The sizable spontaneous leptogenesis in the oscillating phase after inflation is realized by two effective operators: the  Weinberg operator responsible for lepton number violation and the effective ``chemical potential'' operator of the interaction between Higgs-squared and the lepton-number current. While the oscillating Higgs field is being damped out, the time dependent masses induced by the lepton number violating operator would lead to production of neutrinos and anti-neutrinos. The chemical potential makes an asymmetric between the generated lepton and anti-lepton numbers, while maintaining consistency with high reheating temperature in Higgs inflation: $\widetilde T \in (10^{15} - 10^{18})~\text{GeV}$. The sphaleron processes eventually convert the lepton asymmetry into baryon asymmetry. We emphasize that our scenario is a minimal realization of baryogenesis since  we do not require any extra degree of freedom beyond the SM particles. The precise measurement of the mass of top quark ($m_t$), strong coupling constant ($\alpha_s$) and the Higgs self-coupling ($\lambda$) from the collider experiments in the future will solidify or disprove the validity of our scenario.

As a final remark, in critical Higgs inflation \cite{Hamada:2014iga,Bezrukov:2014bra,Hamada:2014wna}, relatively small non-minimal coupling as small as  $ \xi \lsim \mathcal{ O}(10) $ is allowed. In this case,  asymmetry generation becomes more efficient and there are more allowed parameter spaces. In addition, the unitarity problem is relieved and the dangerous violent preheating is tame. We leave analysis of the reheating processes and spontaneous leptogenesis in the critical Higgs inflation for future.

\acknowledgments

We thank Fedor Bezrukov, Kohei Kamada, Min-xi He and Yongsoo Jho for helpful discussions and comments. This work was supported by the National Research Foundation of Korea (NRF) grant funded by the Korea government (MOE) (No.~2020R1A6A3A13076216) (SML) and (MSIT) (NRF2018R1A4A1025334), (NRF-2019R1A2C1089334) (SCP). SML was supported in part by the Hyundai Motor Chung Mong-Koo Foundation. The work of KO is in part supported by JSPS Kakenhi Grant No.~19H01899. 

\appendix
\section{Conformal Transformation} \label{App:ConfTrsf}
In this section, we will review the conformal transformations on inflaton and on a relevant fermion sector. We set $ M_{P} \equiv 1 $ here.
\subsection{Inflaton Sector}
Let us consider the gravity and the non-minimally coupled inflaton action in the Jordan frame:
\begin{align}  \label{Eq:infaction}
S_{\rm inf} = \int d^{4}x \sqrt{-g_{J}} \left[\frac{1}{2} \Omega^{2} R_{J} 	- \frac{1}{2} g_{J}^{\mu\nu} \partial_{\mu} \phi_{J} \partial_{\nu} \phi_{J}  - V_{J}(\phi_{J})	\right],
\end{align}
where we consider a conformal factor
\begin{align} \label{Eq:confac}
\Omega^{2} = 1 + \xi \phi_{J}^{2}.
\end{align}
By the conformal transformation, the metric in the Einstein frame $ g_{\mu\nu} $ is
\begin{align} \label{Eq:metricJtoE}
g_{\mu\nu} = \Omega^{2} g_{J\mu\nu}.
\end{align}
Note that we do not put subscript `$ E $' explicitly. By the conformal transformation, it is well known that the Ricci scalar also transforms to (in the metric formalism):
\begin{align}
R_{J} = \Omega^{2} \left[ R + 3 g^{\mu\nu} \partial_{\mu} \partial_{\nu} \ln \Omega^{2} - \frac{3}{2} g^{\mu\nu} \partial_{\mu} \ln \Omega^{2}  \partial_{\nu} \ln \Omega^{2}  \right].
\end{align}
When we substitute Eq.~(\ref{Eq:confac}) and Eq.~(\ref{Eq:metricJtoE}) to the action Eq.~(\ref{Eq:infaction}) becomes
\begin{align}
	S = \int d^{4}x \sqrt{-g} \left[ R - \frac{1}{2} \left( \frac{\Omega^{2} + 6 \xi^{2} \phi_{J}^{2}}{\Omega^{4}}	\right ) g^{\mu\nu} \partial_{\mu} \phi_{J} \partial_{\nu} \phi_{J}	- \frac{V_{J}(\phi_{J})}{\Omega^{4}}	\right].
\end{align}
To make its kinetic term canonical, we redefine the inflaton field: The Einstein frame field $ \phi $ is given by
\begin{align}
\frac{d\phi}{d \phi_{J}} =  \left( \frac{\Omega^{2} + 6 \xi^{2} \phi_{J}^{2}}{\Omega^{4}}	\right )^{1/2}.
\end{align}
This can be analytically integrated:
\begin{align} 
\phi = \sqrt{\frac{1+6\xi}{\xi}} \sinh^{-1} \left(\sqrt{\xi(1+6\xi)}\phi_{J}	\right) - \sqrt{6} \sinh^{-1} \left(	\sqrt{\frac{6\xi^{2} \phi_{J}^{2} }{1 + \xi \phi_{J}^{2} }}	\right).
\end{align}
In the limit of $ \xi \gg 1 $, this relation can be approximated as\footnote{
Here we only consider the absolute value of $ \phi $ following the logic of \cite{GarciaBellido:2008ab}.}
\begin{align}
\label{eq:phiphiJ}
\phi \simeq \begin{cases}
\phi_{J} & \vert  \phi \vert \ll \frac{1}{\xi} \\
\sqrt{\frac{3}{2}} \ln \left(	1 + \xi \phi_{J}^{2}	\right) & \vert \phi \vert \gg \frac{1}{\xi} 
\end{cases}.
\end{align}

{For the intermediate regime $1/{\xi} \ll  \phi \lsim 1/{\sqrt{\xi}}$, the latter relation in Eq.~\eqref{eq:phiphiJ} implies $\Omega^2 \simeq e^{\sqrt{\frac{2}{3}} \vert \phi \vert}$, and $\xi \phi_J^2 \simeq e^{\sqrt{\frac{2}{3}}\phi} -1 \simeq  \sqrt{\frac{2}{3}} \phi$, that is,
\begin{align}
   \phi \simeq \sqrt{\frac{3}{2}} \xi \phi_{J}^2. 
\end{align}
Then the Jordan-frame potential $ V_{J}(\phi_{J}) = \frac{\lambda}{4}\phi_{J}^{4} $ leads to the Einstein-frame potential $ V_{E} = V_{J}/\Omega^{4} $ 
in the intermediate regime relevant to the reheating phase:
\begin{align}
V_{E}(\phi)
	&=	\frac{\lambda \phi_J^4/4}{ \left(1+\xi \phi_J^2 \right)^{2}}
	\simeq   \frac{\lambda M_{P}^{2}}{6 \xi^{2}} \phi^{2}.
\end{align}
This corresponds to Eq.~(\ref{Eq:quadraticpotential}) with $M^{2}=\frac{\lambda M_{P}^{2}}{3 \xi^{2}}$.
}

\subsection{Fermion Sector}
Under the conformal transformation, fermion sector also changes. We consider a Dirac spinor Langrangian as well as Majorana neutrino with mass $ M_{N} $ and the dimension-six inflaton-current interaction term:
\begin{align}
	S_{f}= \int d^{4}x \sqrt{-g_{J}} \left[	-\frac{1}{2} \overline{\psi}_{J} {e_{Ja}}^{\mu} \gamma^{a} D_{J\mu} \psi_{J} - \frac{1}{2}(M_{N} + y \phi_{J} ) \overline{\psi}_{J} \psi_{J} + \frac{c}{\Lambda^{2}}\left(\partial_{\mu}\phi_{J}^{2}\right) J_{J L}^{\mu}
	\right],
\end{align}
where $ {e_{Ja}}^{\mu} $ is vielbein in the Jordan frame and
\begin{align}
	D_{J\mu} = \partial_{\mu} + \frac{1}{4}\omega_{J a b \mu} \gamma^{[a} \gamma^{b]}
\end{align}
is covariant derivative of spinor with the spin connection $ \omega_{J a b \mu} $
\begin{align}
\omega_{J a b \mu} = e_{Ja\nu} \nabla_{\mu} {e_{Jb}}^{\nu}.
\end{align}
Here the covariant derivative $ \nabla_{\mu} $ is defined by
\begin{align}
	\nabla_{\mu} {e_{Ja}}^{\nu} = \partial_{\mu} {e_{Ja}}^{\nu} + \Gamma^{\nu}_{\lambda \mu}  {e_{Ja}}^{\lambda},
\end{align}
with $ \Gamma^{\nu}_{\lambda \mu} $ being the affine connection. The spin connection in the Jordan frame $ \omega_{J a b \mu} $ is related to the spin connection $ \omega_{J a b} $ and vielbein $ e_{a\mu} $ in the Einstein frame 
\begin{align}
	\omega_{Jab\mu} = \omega_{ab\mu} + e_{b \mu} \partial_{a} \ln \Omega- e_{a \mu} \partial_{b} \ln \Omega
\end{align}
and this results in
\begin{align}
	\gamma^{a} {e_{J a}}^{\mu} D_{J \mu} \psi_{J} = \Omega^{5/2} \gamma^{a}{e_{a}}^{\mu} D_{\mu} \psi,
\end{align}
where we redefine the spinor field in the Einstein frame $ \psi $ by 
\begin{align}
\psi_{J} = \Omega^{3/2}\psi.
\end{align}
Then, the transformation of the current is
\begin{align}
	J_{J}^{\mu} = \overline{\psi}_{J} {e_{Ja}}^{\mu} \gamma^{a} \psi_{J} = \Omega^{4} \overline{\psi} {e_{a}}^{\mu} \gamma^{a} \psi = \Omega^{4} J^{\mu}.
\end{align}
The final form of the fermion sector becomes
\begin{align}
S_{f}= \int d^{4}x \sqrt{-g} \left[	-\frac{1}{2} \overline{\psi} {e_{a}}^{\mu} \gamma^{a} D_{\mu} \psi - \frac{1}{2}\left( \frac{M_{N} +  y \phi_{J}}{\Omega} \right) \overline{\psi} \psi +\frac{c}{\Lambda^{2}} (\partial_{\mu}\phi_{J}^{2}) J_{L}^{\mu}
\right].
\end{align}
Note that the Dirac and Majorana mass terms change differently as
\begin{align}
M_{N} 
	&\rightarrow
		\frac{M_{N}}{\Omega} = M_{N} e^{-\frac{1}{\sqrt{6}}\frac{\vert \phi \vert}{M_{P}}},&
M_{D}
	&\rightarrow
		\frac{y}{\Omega}\phi_{J} = \frac{y M_{P}}{\sqrt{\xi}} \left(	1- e^{-\sqrt{\frac{2}{3}} \frac{\vert \phi \vert}{M_{P}}} \right)^{1/2}.
\end{align}
If the origin of neutrino masses is the seesaw mechanism,
\begin{align}
	m_{\nu}=\frac{M_{D}^{2}}{M_{N}} \rightarrow \frac{y^{2} \phi_{J}^{2}}{M_{N} \Omega} = \frac{2 y^{2}M_{P}^{2}}{M_{N} \xi} \sinh \left(\frac{\vert \phi \vert}{\sqrt{6}}\right).
\end{align}
This result can also be derived from the dimension-five Weinberg operator as in the main text:
\begin{align}
	\sqrt{-g_{J}}\frac{c_{5}}{M_{P}}\phi_{J}^{2} \psi_{J} \psi_{J} = \sqrt{-g}\frac{c_{5}}{M_{P}} \frac{\phi_{J}^{2}}{\Omega} \psi\psi.
\end{align}
with identification $ y^{2}/M_{N} = c_{5}/M_{P} $.

\section{Bogoliubov Transformation} \label{App:BogoTrsf}
This section presents details of Bogoliubov transformation used in Section \ref{Neutrino Production}. The material closely follows Ref.~\cite{Pearce:2015nga}, with minor changes of notations.
In this section, we will mainly use comoving coordinates: the conformal time $ \tau \equiv \int dt /a 	 $ and the comoving momentum $ k \equiv a p $, with $ a $ and $p$ being the scale factor and the physical momentum, respectively. We also define $ \widetilde{m}_{\nu} \equiv a m_{\nu}$ and $\widetilde{\mu} \equiv a \mu $. We will denote $ F^{\prime} \equiv \partial_{\tau} F $ and $ \dot{F} \equiv \partial_{t} F$ for any function $ F $.

The equation of motion for neutrino with additional term Eq.~(\ref{Eq:chemical potential operator}) is
\begin{align} \label{appeq:eom}
( i \partial_{\tau}  - i \vec{\sigma} \cdot \vec{\partial} ) \nu_{L} = -\widetilde{m}_{\nu} (i \sigma_{2}) \nu_{L}^{*} - \widetilde{\mu} \nu_{L},
\end{align}
or in the momentum space
\begin{align}
( i \partial_{\tau} + \vec{\sigma} \cdot \vec{k} ) \nu_{L} = - \widetilde{m}_{\nu} (i \sigma_{2}) \nu_{L}^{*} - \widetilde{\mu} \nu_{L}.
\end{align}

Mode expansion of the neutrino field is
\begin{align}
\nu_{L} = \int \frac{d^{3}k}{(2 \pi)^{3}} \sum_{s=\pm 1} \left[\xi_{s}(\tau,\vec{k}) a_{s}(\vec{k}) e^{i \vec{k} \cdot \vec{x}} + 
\chi_{s}(\tau,\vec{k}) a_{s}^{\dagger}(\vec{k}) e^{-i \vec{k} \cdot \vec{x}}	\right],
\end{align}
where $ s  $ is the helicity index;
$ a_{s} $ and $ a_{s}^{\dagger} $ are annihilation and creation operators, respectively, satisfying the anticommutation relation
\begin{align}
\left\{ a_{s}(\vec{k}), a_{s^{\prime}}^{\dagger}(\vec{k}^{\prime}) \right\} = (2 \pi)^{3} \delta_{s s^{\prime}} \delta(\vec{k} - \vec{k}^{\prime});
\end{align}
$ \xi_{s} $ and $ \chi_{s} $ are coefficients with definite helicities and spinor indices are suppressed.

Using the unit helicity eigenstate $ h_{s} $ such that
\begin{align}
\hat{h} h_{s} (\vec{k}) = s  h_{s} (\vec{k}),&&    \hat{h} \equiv	\vec{S} \cdot \hat{k}= \frac{1}{2} \vec{\sigma} \cdot \hat{k} 
\end{align}
where $ \hat{k} = \vec{k}/ \vert \vec{k}\vert$
and
\begin{align}
h_{s}^{\dagger}(\vec{k}) h_{s^{\prime}}(\vec{k}) = \delta_{ss^{\prime}},&&\sum_{s=\pm 1}h_{s}(\vec{k}) h_{s}^{\dagger}(\vec{k}) = 1,
\end{align}
we rewrite coefficients $ \xi_{s} $ and $ \chi_{s} $ as

\begin{align}
\xi_{s} (\tau, \vec{k}) = u_{s}(\tau,\vec{k}) h_{s}(\vec{k}),&&
\chi_{s}^{c} (\tau, \vec{k}) = v_{s}(\tau,\vec{k}) h_{s}(\vec{k})
\end{align}
where $ \chi^{c} \equiv i \sigma_{2} \chi^{*} $ in our notation. We choose $ \chi^{c} $ so that $ u $ and $ v $ both correspond to positive frequency modes.

Substituting this mode expansion to Eq.~(\ref{appeq:eom}), we obtain coupled differential equations for $ u_{h} $ and $ v_{h} $
\begin{align}\label{Eq:original}
(i \partial_{\tau} + s k ) u_{s} = - \widetilde{m}_{\nu} v_{s} - \widetilde{\mu} u_{s},&&	( i \partial_{\tau} - s k ) v_{s} = -\widetilde{m}_{\nu} u_{s} + \widetilde{\mu} v_{s}.
\end{align}
For the simplicity of the calculation, we perform additional expansion,
\begin{align} \label{appEq:modeexp}
u_{s}(\tau,\vec{k}) &= \frac{\alpha_{s}}{\sqrt{2}}\sqrt{1-f_{s}} e^{-i   \int^{\tau}_{0} \omega_{s}(\tau^{\prime}) d\tau^{\prime}} + \frac{\beta_{s}}{\sqrt{2}} \sqrt{1+f_{s}} e^{i  \int^{\tau}_{0} \omega_{s}(\tau^{\prime}) d\tau^{\prime}}, \nonumber\\
v_{s}(\tau,\vec{k})  &= - \frac{ \alpha_{s}}{\sqrt{2}}\sqrt{1+f_{s}} e^{-i \int^{\tau}_{0} \omega_{s}(\tau^{\prime}) d\tau^{\prime}} + \frac{ \beta_{s}}{\sqrt{2}} \sqrt{1- f_{s}} e^{i  \int^{\tau}_{0} \omega_{s}(\tau^{\prime}) d\tau^{\prime}},
\end{align}
where
\begin{align}\label{appeq:uandv}
\omega_{s}^{2}(\tau,\vec{k})
&=
(k + s \widetilde{\mu})^{2} + \widetilde{m}_{\nu}^{2},&
f_{s}
&\equiv
\frac{sk + \widetilde{\mu}}{\omega_{s}}.
\end{align}
Here one can see explicit difference between of the dispersion relations between two helicities due to the existence of $ \mu $ and this eventually becomes the source of the lepton asymmetry.

Just for the simplicity of the notation, We will omit explicit $ s $ subscript for a while. In addition, we will consider $ \omega $ as a constant so that
$ 	\int^{\tau}_{0} d \tau^{\prime} \omega(\tau^{\prime}) = \omega \tau $.
We emphasize that this does not ruin the generality of the following derivations. We also denote $ g_{\pm} \equiv \sqrt{1 \pm f} $. Then, the the Eq.~(\ref{appEq:modeexp}) becomes
\begin{align}
u(\tau,\vec{k}) &= \frac{\alpha}{\sqrt{2}}
g_{-} e^{-i \omega \tau} + \frac{\beta}{\sqrt{2}} g_{+} e^{i  \omega \tau} \nonumber\\
v(\tau,\vec{k})  &= - \frac{ \alpha}{\sqrt{2}}g_{+} e^{-i \omega \tau} + \frac{ \beta}{\sqrt{2}} g_{-} e^{i  \omega \tau}.
\end{align}
From the definition of $ f $,
\begin{align}
f^{2} = \frac{(k + s\widetilde{\mu})^{2}}{(k + s\widetilde{\mu})^{2} + \widetilde{m}_{\nu}^{2}},&&g_{+}g_{-} = \sqrt{1-f^{2}} = \frac{m_{\nu}}{\omega}.
\end{align}
The first equation of Eq.~(\ref{Eq:original}) for $ u $ and $ v $  can be rearranged to
\begin{align}
(i \partial_{\tau} + f \omega ) u = - (g_{+}g_{-}) \omega v.
\end{align}
Plugging Eq.~(\ref{appeq:uandv}) into this equation gives
\begin{align}
&(\alpha^{\prime} g_{-} + \alpha g_{-}^{\prime} )  e^{-i \omega \tau} +	(\beta^{\prime} g_{+}+\beta g_{+}^{\prime})  e^{i \omega \tau} \nonumber\\&~~~+ \alpha  g_{-} (1+f)\omega e^{-i\omega\tau}-\beta
g_{+} (1-f)\omega e^{i\omega\tau} = - \sqrt{2}(g_{+}g_{-})\omega v 
\end{align}
Note that the terms in the second line exactly cancelled as
\begin{align}
&\alpha  g_{-} (1+f)\omega e^{-i\omega\tau}-\beta
g_{+} (1-f)\omega e^{i\omega\tau} \nonumber\\&= g_{+}g_{-} \omega \left( \alpha  g_{+}\omega e^{-i\omega\tau}-\beta
g_{-} e^{i\omega\tau}			\right) = - \sqrt{2}(g_{+}g_{-})\omega v,
\end{align}
and the similar procedure can be applied to the second equation of Eq.~(\ref{Eq:original}).

Finally, we obtain a set of equations
\begin{align}
&(1)~~~\alpha^{\prime} g_{-}  e^{-i\int \omega d\tau} +	\beta^{\prime} g_{+}  e^{i\int \omega d\tau} + \alpha g_{-}^{\prime} e^{-i \int \omega d\tau} + \beta g_{+}^{\prime} e^{+i\int \omega d\tau} =0 \nonumber\\
&(2)~~~\alpha^{\prime} g_{+}  e^{-i\int \omega d\tau} -	\beta^{\prime} g_{-}  e^{i\int \omega d\tau} + \alpha g_{+}^{\prime} e^{-i \int \omega d\tau} - \beta g_{-}^{\prime} e^{+i \int \omega d\tau} = 0.
\end{align}
By multiplying $ g_{+} $ and $ g_{-} $,
\begin{align}
&\left( (1)\times g_{-} + (2) \times g_{+} \right): \nonumber\\&~~~\alpha^{\prime} (g_{-}^{2} + g_{+}^{2}) e^{-i \omega \tau} = - \alpha (g_{-} g_{-}^{\prime}+g_{+} g_{+}^{\prime}) e^{-i \omega \tau} - \beta(g_{-} g_{+}^{\prime} - g_{+} g_{-}^{\prime})e^{i \omega \tau} \nonumber\\
&\left( (1)\times g_{+} - (2) \times g_{-} \right): \nonumber\\&~~~\beta^{\prime} (g_{-}^{2} + g_{+}^{2}) e^{i \omega \tau} = - \alpha (g_{+} g_{-}^{\prime} - g_{-} g_{+}^{\prime}) e^{-i \omega \tau} - \beta(g_{+} g_{+}^{\prime} + g_{-} g_{-}^{\prime})e^{i \omega \tau}. \label{Eq:App.complex}
\end{align}
One can easily prove that there are useful relations between $ g_{+} $ and $ g_{-} $ as followings:
\begin{align}
g_{+}^{2} + g_{-}^{2} = 2,&&
g_{-}g_{-}^{\prime} + g_{+}g_{+}^{\prime} = 0,&&
g_{-}g_{+}^{\prime} - g_{+}g_{-}^{\prime} = \frac{f^{\prime}}{\sqrt{1-f^{2}}}.
\end{align}
Then, Eq.~(\ref{Eq:App.complex}) can be simplified as
\begin{align}
\alpha^{\prime} =  - \frac{\beta}{2} \frac{f^{\prime}}{\sqrt{1-f^{2}}} e^{2 i \omega \tau},&&\beta^{\prime} =    \frac{\alpha}{2} \frac{f^{\prime}}{\sqrt{1-f^{2}}} e^{-2 i \omega \tau},
\end{align}
which leads us to the desired result corresponding to Eq.~(\ref{Eq:coupled}) after reintroducting explicit $ s $ index and integrals:
\begin{align} \label{appEq:coupled}
\alpha_{s}^{\prime}(\tau,k) = - \frac{\beta_{s}(\tau,k)}{2\omega_{s}^{2}} \left[	\widetilde{m}_{\nu} \widetilde{\mu}^{\prime} - (s k + \widetilde{\mu}) \widetilde{m}_{\nu}^{\prime}	\right] e^{2 i \int^{\tau}_{0} \omega_{s}(\tau^{\prime}) d\tau^{\prime}}, \nonumber\\
\beta_{s}^{\prime}(\tau,k) =  \frac{\alpha_{s}(\tau,k)}{2\omega_{s}^{2}} \left[	\widetilde{m}_{\nu} \widetilde{\mu}^{\prime} - (s k + \widetilde{\mu}) \widetilde{m}_{\nu}^{\prime}	\right] e^{-2 i \int^{\tau}_{0} \omega_{s}(\tau^{\prime}) d\tau^{\prime} },
\end{align}
with the initial conditions $ \alpha_{s}(0,k) = 1 $ and $ \beta_{s}(0,k)=0 $, where we take $ \tau=0 $ at the end of the inflation. This corresponds to choosing the Bunch-Davies vacuum (having only positive frequencies) for the initial state of neutrino field at the end of inflation~\cite{Birrell:1982ix,Mukhanov:2007zz}.

One can further introduce new variables $ A $ and $ B $ for numerical convenience as
\begin{align}
A_{s}(\tau,k) &\equiv \alpha_{s}(\tau,k) e^{-i \int^{\tau}_{0} \omega_{s}(\tau^{\prime}) d\tau^{\prime}},&
B_{s}(\tau,k) &\equiv \beta_{s}(\tau,k) e^{i \int^{\tau}_{0} \omega_{s}(\tau^{\prime}) d\tau^{\prime}},
\end{align}
to make Eq.~(\ref{appEq:coupled})
\begin{align}
A_{s}^{\prime} (\tau,k)
&= - c_{s} B_{s}(\tau,k)  - i \omega_{s} A_{k}(\tau,k),&
B_{k}^{\prime}(\tau,k) 
&=  c_{s} A_{s}(\tau,k)  +  i \omega_{s} B_{s}(\tau,k),
\end{align}
where
\begin{align}
c_{s}(\tau,k) \equiv \frac{1}{2\omega_{s}^{2}} \left[	\widetilde{m}_{\nu} \widetilde{\mu}^{\prime} - (sk + \widetilde{\mu}) \widetilde{m}_{\nu}^{\prime}	\right] = \frac{s}{2\omega^{2}} \left[	\widetilde{m}_{\nu} (s\widetilde{\mu})^{\prime} - (k + s\widetilde{\mu})\widetilde{m}_{\nu}^{\prime}	\right],
\end{align}
with the initial conditions $A_{s}(0,k) = 1 $ and $ B_{s}(0,k)=0 $. 

We will solve the equation in the physical Einstein-frame time and the physical momentum,
\begin{align} \label{EqApp:DiffEq}
\dot{A}_{s}(t,p) 
&= - C_{s} B_{s}(t,p) - i \omega_{s} A_{s}(t,p), &
\dot{B}_{s}(t,p) 
&=  C_{s} A_{s}(t,p) + i \omega_{s} B_{s}(t,p),
\end{align}
with
\begin{align} \label{EqApp:CDef}
C_{s}(t,p) &= \frac{s}{2\omega^{2}} \left[m_{\nu} (H s\mu + s\dot{\mu})	 - (p + s \mu )(H m_{\nu}  + \dot{m}_{\nu})\right] \nonumber \\
&\simeq \frac{s}{2\omega^{2}} \left[m_{\nu} s \dot{\mu}	 - (p + s \mu ) \dot{m}_{\nu}\right],
\end{align}
where we neglect terms proportional to Hubble parameter $ H(t) $ in the last step by the fact that frequency of $ \mu $ and $ m_{\nu} $ is much larger than the Hubble scale during the reheating.

\section{Relation between $ T_{\rm reh} $, $ N_{\rm reh} $, and $ n_{s} $} 
In this section, we will the derivations of the relation between $ T_{\rm reh} $, $ N_{\rm reh} $, and $ n_{s} $ following Ref.~\cite{Cook:2015vqa}. Similar calculation has been conducted by \cite{Cai:2015soa, Gong:2015qha}.

\label{App:RelationTemp}
Tracing back from the current temperature of universe $ T_{0} \simeq 2.73 ~\rm K $, we can deduce the reheating temperature
\begin{align}
T_{\rm reh} &= T_{0} \left(\frac{a_{0}}{a_{\rm reh}}	\right)\left(	\frac{43}{11 g_{*,\rm reh}}	\right)^{\frac{1}{3}}
\\& = T_{0} \left( \frac{a_{0}}{a_{\rm eq}}\right) 
\left(	\frac{a_{\rm eq}}{a_{\rm reh}}	\right) \left(	\frac{43}{11 g_{*,\rm reh}}	\right)^{\frac{1}{3}} = T_{0} \left( \frac{a_{0}}{a_{\rm eq}}\right) 
e^{N_{\rm rad}}\left(	\frac{43}{11 g_{*,\rm reh}}	\right)^{\frac{1}{3}}
\end{align}
where $ a_{0} $, $ a_{\rm eq} $, $ a_{\rm reh} $ is the scale factor at the time of the end of the inflation, matter-radition equality, and at the end of reheating, respectively. For future use, we will denote the scale factor when CMB pivot scale $ (k_{*}/a_{0} = 0.05~\rm Mpc^{-1}) $ leaves the horizon as $ a_{*} $. The $ g_{*,\rm reh} $ is the effective relativistic degrees of freedom at the end of the reheating. We will take $ g_{*,\rm reh} = 106.75 $.

By defining e-folding numbers for each stages during the history of the universe as
\begin{align}
	N_{*} = \ln  \left( \frac{a_{\rm end}}{a_{*}}\right), &&
	N_{\rm reh} = \ln  \left( \frac{a_{\rm reh}}{a_{\rm end}}\right),&&
	N_{\rm rad} = \ln  \left( \frac{a_{\rm eq}}{a_{\rm reh}}\right)
\end{align}
we have following relation:
\begin{align}
\frac{a_{0}}{a_{\rm eq}} 
=\frac{a_{0}}{a_{*}} \frac{a_{*}}{a_{\rm end}} \frac{a_{\rm end}}{a_{\rm reh}} \frac{a_{\rm reh}}{a_{\rm eq}}
= \frac{a_{0}H_{*}}{k_{*}} e^{-N_{*}} e^{-N_{\rm reh}} e^{-N_{\rm rad}}.
\end{align}
where we used $ k_{*} = a_{*} H_{*} $ at last equality.

Therefore, rescaled temperature $ \widetilde{T} \equiv e^{N_{\rm reh}}T_{\rm reh} $ can be determined by the Hubble scale at pivot scale $ H_{*} $ and e-folding number between time when the pivot scale $ k_{*} $ crosses the horizon and the time at the end of the inflation $ N_{*} $,
\begin{align}
\widetilde{T} = e^{N_{\rm reh}}T_{\rm reh} = \left(	\frac{43}{11 g_{*,\rm reh}}	\right)^{\frac{1}{3}}
\left(\frac{a_{0} T_{0}}{k_{*}}\right) H_{*} e^{-N_{*}}.
\end{align}

For a given model of inflation, $ H_{*} $ and $ N_{*} $ are generally related to the slow-roll parameters $ \epsilon $ and $ \eta $, hence the observables $ n_{s} $ and $ r $ at pivot scale. In the case of Higgs inflation \cite{Bezrukov:2007ep},
\begin{align}
	\epsilon_{*} \simeq \frac{3}{4 N_{*}^{2}},&&\eta_{*} \simeq -\frac{1}{N_{*}},&& n_{s} \simeq 1 - 6 \epsilon_{*} + 2 \eta_{*} \simeq 1- \frac{2}{N_{*}},&&N_{*} = \frac{2}{1-n_{s}}. 
\end{align}
The Hubble scale also determines the amplitude of the power spectrum is given by
\begin{align}
	A_{s} = \frac{1}{8 \pi^{2} \epsilon_{*}} \frac{H_{*}^{2}}{M_{P}^{2}},&&H_{*} = \pi M_{P} \sqrt{\frac{3}{2}A_{s}} (1-n_{s}).
\end{align}
Hence, we can evaluate $ \widetilde{T}$ solely as a function of $ n_{s} $. Note that we does not assume any equation of state during the reheating so far.

$ T_{\rm reh} $ and $ N_{\rm reh} $ are determined seperately when specific equation of state is given \cite{Cook:2015vqa}:
\begin{align}
	T_{\rm reh} &= \left[\left(\frac{43}{11g_{*,\rm reh}}\right)^{\frac{1}{3}} \frac{a_{0}T_{0}}{k_{*}} H_{*} e^{-N_{*}} \left(\frac{45 V_{\rm end}}{\pi^{2} g_{*,\rm reh}}\right)^{- \frac{1}{3(1+w)}}			\right]^{\frac{3(1+w)}{3w - 1}}\\
	N_{\rm reh}&= \frac{4}{1-3w} \left[ - \frac{1}{4} \ln \left(	\frac{45}{\pi^{2} g_{*,\rm reh}} \right) - \frac{1}{3} \ln \left(	\frac{11g_{*,\rm reh}}{43}\right) - \ln\left(\frac{k_{*}}{a_{0} T_{0}}\right) - \ln \left(\frac{V_{\rm end}^{1/4}}{H_{*}} \right) - N_{*}  \right] \\
	&= \frac{4}{1-3w}   \left[  61.6 - \ln \left(\frac{V_{\rm end}^{1/4}}{H_{*}} \right) - N_{*}  \right]
\end{align}
where
\begin{align}
	V_{\rm end} = \frac{9}{2} \pi^{2} M_{P}^{4} A_{s} (1-n_{s})^{2} \frac{256}{(2+\sqrt{3})^{2} (5+3n_{s})^{2} }
\end{align}
for Higgs inflation.

\bibliographystyle{utphys}
\bibliography{leptogenesis.bib}

\end{document}